# Effects of the COVID-19 pandemic in higher education: a particular case from the perspective of complex systems

**Fátima Velásquez-Rojas**, *Instituto de Física de Líquidos y Sistemas Biológicos (UNLP-CONICET) & Departamento de Ciencias Básicas, Facultad de Ingeniería, Universidad Nacional de La Plata (UNLP), B1900BTE La Plata, Argentina.*
**Jesús E. Fajardo**, *Departamento de Física Médica, Centro Atómico Bariloche, CONICET, CNEA, R8402AGP Bariloche, Argentina.*
**Daniela Zacharias**, *Departamento de Estadística, Universidad Nacional del Comahue (UNCOMA), R8402AGP Bariloche, Argentina.*
**María Fabiana Laguna**, *Centro Atómico Bariloche, Consejo Nacional de Investigaciones Científicas y Técnicas (CONICET) & Profesorado en Física, Universidad Nacional de Río Negro (UNRN), R8402AGP Bariloche, Argentina.*

**Abstract**
The COVID-19 pandemic abruptly changed the classroom context, understood as the meeting space between teachers and students where a fundamental part of the construction of new knowledge takes place. This presented enormous challenges for all actors in the educational process, who had to overcome multiple difficulties and adapt to a new daily life in which the incorporation of new strategies and tools was essential. In this work we study the knowledge acquisition process in two different contexts: face-to-face (before the onset of the pandemic) and virtual (during confinement), for a particular case in higher education in Argentina. We developed an analytical model for the knowledge acquisition process, based on a series of surveys and information on academic performance. We analyzed the significance of the model by means of Artificial Neural Networks and a Multiple Linear Regression Method. We found that the virtual context produced a decrease in motivation to learn. We also found that the structure of the emerging contact network from peer interaction presents very different characteristics in both contexts. Finally, we show that in all cases, interaction with teachers is of utmost importance in the process of acquiring knowledge.

**Introduction**

One of the most effective strategies to mitigate a pandemic is isolation [1-9]. This measure, taken by many countries to mitigate the effects of COVID-19, changed our way of life and involved modifying the way we educate and are educated. The lockdowns in response to COVID-19 have disrupted conventional education with school and university closures around the world. While the educational community has made efforts to maintain learning continuity during this period, students have had to rely more on their own resources to continue learning remotely [10].

In Argentina, the confinement measures that affected the educational level were carried out from March 20, 2020 and this coincided with the beginning of the first semester of the academic year. This unconventional schooling context led to great challenges and was reflected in the results obtained by the students [11-15]. The situation has been and has been analyzed from different perspectives, but one that we are interested in highlighting is the analysis of the knowledge acquisition process (KAP) from the point of view of complex systems, in line with what has been done by some of the authors of the present work [16]. In that work we developed an analytical model (the KA model) based on data from a series of surveys that are contrasted with information on academic performance of students, to analyze how the KAP depends globally on different extrinsic and intrinsic factors.

Regarding the intrinsic factors, one that contributes greatly to the acquisition of knowledge of students is motivation, and this is precisely one of the most affected by the pandemic [17]. According to the EU report [11], the closure of physical schools and the adoption of distance education can negatively affect student learning through four main channels: less time spent learning, symptoms of stress, a change in the way that students interact and lack of motivation to learn. However, despite this, distance education is essential to ensure the continuity of learning in situations in which face-to-face classes are suspended.

On the other hand, it was already mentioned that the change in physical context affected extrinsic factors that contribute to the acquisition of knowledge, such as the interaction with peers and teachers. This interaction has been found to be essential for the development of positive self-esteem, self-confidence, and a sense of identity. In fact, there is significant evidence showing that social skills are positively associated with cognitive skills and school achievement [18,19].

In this new approach we adapt the analytical model presented in [16] to compare the KAP in two different contexts: face-to-face (before the onset of the pandemic) and virtual (during the confinement), for a particular case in higher education in Argentina. Besides, and in order to evaluate the relevance of the parameters that we choose for our model, we used Artificial Neural Networks and a Multiple Linear Regression Method.

The article is organized as follows: in the Methods section we describe the participants and its educational context, the measures (which include the surveys used to construct our data-based model) and the different approaches used to fit the parameters of the model. Then, we present the main results of this work and finally, we summarize and discuss our findings.

**Methods**

**Participants**

The research was carried out with several sections of students who attended the Physics II course, corresponding to the second year of Engineering careers at the Faculty of Engineering of the National University of La Plata (UNLP) [20], Argentina, during the years 2020 and 2021. The Faculty offers 13 engineering degrees, so the interest of the students in the course can vary greatly.

The complete course lasts one semester, with a workload of 8 hours per week divided into 2 theoretical-practical classes. The course consists of two parts, at the end of which a partial written test is taken with a score between 0 and 10. There are two approval regimes: direct promotion, which implies being exempt from the final test (if the average between the two partial exams is 6 or more) or promotion by final exam (if the average is between 4 and 6). Partial tests have an instance of recuperation during the semester and another at the end of it, where the student can improve any of the lower scores obtained in previous tests. This organization was also maintained during the confinement (in virtual context).

The first part of the research was done during the two semesters of the year 2019, with four different sections in face-to-face context for a total of 81 students (50 male, 31 female). The second part was developed during the year 2020 and also involved four different sections in two semesters, for a total of 92 students (61 male, 31 female). In all cases we had access to the final grade they obtained in the course. In both contexts, we worked with 4 different sections of students for a total of 8 sections, 173 students in 2 years.

We are interested in analyzing and comparing the processes observed in both contexts in terms of the KA model presented in [16]. A first step consisted in carrying out a classification such as that proposed by Bordogna and Albano [21] and which proved to be useful in our previous work. This involved separating the students into three different groups according to their final achievements Kf, which we relate to the final grade obtained in the course. This was done as follows: (a) High-achieving (HA) students: $8 \leq K_f \leq 10$, (b) Average-achieving (AA) students: $6 < K_f < 8$ and (c) Low-achieving (LA) students: $K_f \leq 6$. It is worth noting that students with a final grade lower than 4 are not included in this study.

In Table 1 we show the number of students who participated in the work divided according to their final achievements Kf, that we relate to the final grade obtained in the course. Interestingly, and as we saw in [16], when analyzing the behavior of these groups we note that each of them has qualitatively different characteristics regarding the relevance of the factors considered in the construction of the new knowledge, as it will be clear shortly.

**Table1**
Number of students participating in the work, divided according to their final achievements Kf in both contexts.

| Context | HA | AA | LA | Total |
|---|---|---|---|---|
| Face-to-face | 13 | 34 | 34 | 81 |
| Virtual | 19 | 42 | 31 | 92 |
| **Total** | **32** | **76** | **65** | **173** |

In Fig. 1 we display the final grades obtained for all students that we include in the present work. In filled symbols we plot the data in the face-to-face context and the empty symbols represent the data in the virtual context. These data provide us with the information to contrast our theoretical model. A first look at this graph reveals that the marks obtained in the two contexts were different for the HA and LA groups, while the AA group did not present differences. HA students, whose grades were higher than 8, had on average a better performance in virtual context than in face-to-face context. The opposite is seen with the low-achievement students, LA. To analyze the possible causes of these differences is the objective of the present paper.

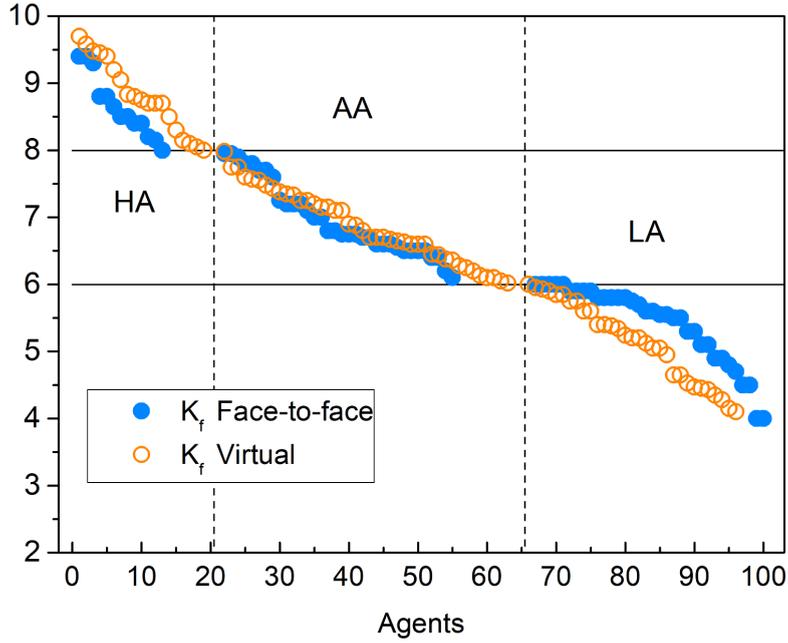

Fig. 1. Group classification of students into three different sets according to their final grade as High-achieving (HA) students, Average-achieving (AA) students and Low-achieving (LA) students for N=173 (filled symbols represent face-to-face context and empty symbols the virtual context). The vertical lines separate the three datasets, while the horizontal lines indicate the scores that we used for the classification made. Data are shown in descending numerical values.

**Measures**

Inspired by our previous work and in order to explicitly compare the classroom observations and surveys with the final achievement of the students, we write the final knowledge on a given topic as the sum of a set of contributing factors. For a student i = 1,2,3, ... N, its final knowledge $K^i_f$ can be posed as:

$$K^i_f = \beta_M^X M^i + \beta_T^X T^i + \beta_P^X P^i \qquad (1)$$

where each factor of Eq. (1) was collected through surveys (of own authorship) in which we analyze their levels of: personal motivation (M) with questions to know experiences and expectations in relation to the course; influence of the teachers (T) and the influence of peers (P). In the next section we detail the numerical values assigned to each factor.

The coefficients $\beta_M^X$, $\beta_T^X$ and $\beta_P^X$ modulates each factor of Eq. (1) and they are different according to the group of students being described, X = HA, AA or LA. Their values were calculated using two methods that we describe further down and that consist of neural networks and a linear regression analysis.

The values of $K^i_f$ obtained from Eq. (1) were compared with the final score obtained by the students, which ranges from 0 to 10.

It is worth noting that in our study the contribution of peers to the acquisition of knowledge was gathered in two ways: the group conformation and the peer interaction itself. The group conformation includes information on the spatial distribution of the students and the formation of groups, obtained through direct observations of the classes before confinement and through questions in online surveys during confinement. An analysis of the differences in the structure of the peer network formed in each context is carried out in Fig. 4 in the Results section.

During the virtual context, important and complementary information was also collected, such as resources the students had (work-space, technological equipment) and the context itself and how it was perceived. Although they are not included as terms in Eq. (1), we carry out a description of the observed situation in the Supporting information S1 File.

Finally, it should be noted that in our study we focus on a specific type of learning, related to scientific concepts of classical physics. While we are aware that this is not the only value learned in the classroom, we simplify the concept of knowledge to use the final grade as a concrete and quantifiable measure of the student's performance.

**Surveys**

The following are the surveys carried out on students during each semester of classes (Table 2). The numbers and letters in the last column correspond to the values that we assign to each of them, in order to transfer the answers to the KA model of Eq. (1). The questions marked with (*) were reformulated to adapt them to the virtual context. The surveys carried out in the virtual context were delivered and completed in a digital way using Google tools, while those corresponding to the pre-confinement stage were delivered personally and were completed manually.

Table 2.
Surveys carried out on students during the semester of classes for two years (the duration of this study). The surveys were broader but here we put the factors involved in the studied model (M, T and P).

| Survey | Quantities | Item | Options | Values |
|---|---|---|---|---|
| **1** (first day of the course) | $M$ | At the beginning of Physics II, what is your level of expectation? | Much / Intermediate / None | 1 / 0,5 / 0 |
| | | Because the course: | Excites me / It is a requirement | 1 / 0 |
| **2** (end of the first part of the course) | $M$ | So far, describe your experience in Physics II: | I really like it / I like it / It is indifferent to me / I do not like it | 1 / 0,5 / 0 / -0,5 |
| | $P$ | At a general level, describe your way of studying: | Alone / In group | 0 / 0,5 |
| **3 (end of the course)** | *M* | At a general level, describe your experience in Physics II: | I really liked it<br>I liked it<br>I was indifferent to me<br>I did not like it | 1<br>0,5<br>0<br>-0,5 |
| | *T* | Were the lectures useful for you? | Yes<br>Little<br>No | 1<br>0,5<br>0 |
| | | Was the interaction with the rest of the teaching team useful to you? | Yes<br>Little<br>No | 1<br>0,5<br>0 |
| | *T\** | Were the virtual lectures useful for you? | Yes<br>Little<br>No | 1<br>0,5<br>0 |
| | | Was the interaction with the rest of the teaching team useful to you? | Yes<br>Little<br>No | 1<br>0,5<br>0 |
| | *T* | At the time of study, which activity was the most beneficial for you? (You can check several options) | Lectures<br>Consulting hours with my group<br>Office hours<br>Private tutoring | A<br>B<br>C<br>D |
| | *T\** | At the time of study, which activity was the most beneficial for you? (You can check several options) | Virtual Lectures<br>Consulting virtual hours with my group<br>Office virtual hours<br>None of the above | A<br>B<br>C<br>D |

The quantities evaluated more than once (as is the case of M or T) were averaged in order to have a single value for each factor. Besides, the combination of strategies for the question that measures T in the third survey was given the following numerical values: ABC, AB, AC, BC=1; A, B=0,7; C, AD, BD, ABD, ACD, BCD = 0,5; CD=0,3; D=0,1 (they could mark several options). These values were given to encourage the use of the strategies provided by the specific section to which the students belonged (options A, B). It is worth remembering that this quantity represents the strategy or strategies to carry out the interaction with the teacher.

As it was already mentioned, each of these groups has different characteristics regarding the relevance of the factors considered in the construction of Eq. (1). To explicitly measure the weight of each of them we apply two different and complementary approaches: Artificial Neural Networks (ANN) and a Multiple Linear Regression Method (MLR).

In reference [22], it has been shown the capability of the single-layer perceptron (SLP) ANN for estimating parameters of complex nonlinear and linear problems, as is our case and the MLR is the most common form of linear regression analysis to treat this kind of problem.

**Single Layer Perceptron (SLP) network overview**
To reproduce Eq.(1) from an ANN architecture we employed a SLP [22]. This type of ANN constitutes a particular case of a Multilayer Perceptron (MLP) [23]. The SLP is a feedforward network of a single artificial neuron-like unit, whose $x_j$ inputs (disposed akin to biologic dendrites) are multiplied by a corresponding weight $w_j$ and this product is passed to a neuron-like unit where the aforementioned product is added up, as shown in Fig. 2.

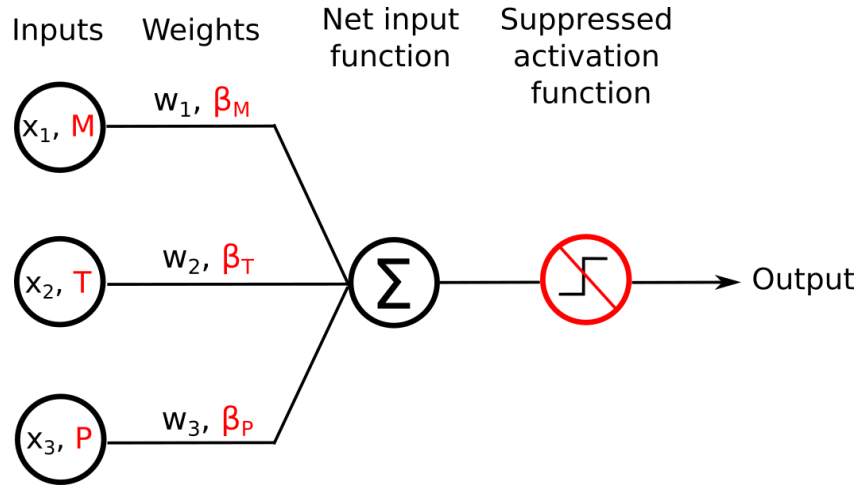

Fig. 2: SLP ANN schematics. The usual ANN notation is in black text and in red text, the equivalent terms corresponding to this particular work are shown. (See Eq. (1)). The elementwise product between the inputs and the weights are added up in the "net input function" stage and suppressing the activation function, an output corresponding to the linear combination of the inputs and the weights is obtained.

This product is typically passed to an activation function to mimic the activation of biological neurons, but this operation can be suppressed by getting the linear combination of the inputs and its weights directly, obtaining the neuron output:

$$z = \sum_{j=1}^{m} x_j w_j = \mathbf{w}^T \mathbf{x}$$

(2)

where Eq. (2) corresponds to a linear combination analogous to Eq. (1).

The inputs $x_j$ and weights $w_j$ in Eq. (2) correspond, respectively to the M, T, P and the coefficients $\beta_M^X, \beta_T^X$ and $\beta_P^X$ in Eq. (1).

The weights (coefficients) values are obtained during the network training. In this stage the experimental data ($M^i$, $T^i$ and $P^i$ of each student i) are passed to the SLP with randomly initialized weights and an output $z_i'$ is calculated. Then the result is compared in each iteration to the actual $z_i$ values (corresponding to each experimental $K_f^i$ of Eq. (1)) and a loss is calculated. As loss metric, we used the mean squared error (MSE) which is one of the commonly used in ANN, particularly in regression applications: $MSE = 1/n \sum_{i=1}^{n}(z_i' - z_i)^2$.

After evaluating the loss, we used the gradient descent (GD) technique to update the gradients. In short, the weights are updated through the network from z to x (right to left in Fig. 2) in all the neuron-like unit in a direction opposite to that of the gradients with the rule:
**w:=w+Δw**
where $\Delta w_j = -\eta \cdot \partial J(w)/\partial w_j$, J(w) is the objective or loss function parameterized to the model parameters **w** and $\eta$ is the learning rate, a parameter chosen between 0 and 1 (typically 0.1 is used for most applications). Once a minimum (ideally a global one) of the function J(w) has been reached and the loss is low enough, the weights vector **w** (corresponding to $\beta_M$, $\beta_P$ and $\beta_T$ coefficients in this particular case) is saved and the multivariate linear model is obtained.

The SLP model was implemented in the programming language Python via the Keras package [24].

**Multiple Linear Regression Method**
Eq. (1) can also be approached by posing a multiple linear regression model (MLR), where the dependent variable is $K_f^i$ and can be explained by the independent variables: M, T and P, taking into account the different groups of students (HA, AA or LA) and the context (face-to-face or virtual) to which each student belongs. Through the MLR method, the measures of the strength of the relationship between the target and predictor variables, the construction of tests of hypothesis and confidence intervals related to regression parameters are expected to be obtained. The model is expressed as:

$$K_f^i = \beta_M M^i + \beta_T T^i + \beta_P P^i + \beta_{HA} HA^i + \beta_{LA} LA^i + \beta_F F^i + \varepsilon^i \quad (3)$$

where i goes from 1 to 173 (the total number of students) and HA, LA are dummy variables created to indicate the group to which the student belongs: HA = 1 if the student belongs to the HA group and HA = 0 if it does not belong to such group. In the same way, LA = 1 if the student belongs to the LA group and LA = 0 if it does not belong to the LA group. AA is considered the reference group, i.e., if a student belongs to this group, HA = 0 and LA = 0. Moreover, F is a dummy variable created to indicate the specific physical context to which the student belongs: F = 1 if the student participated in the face-to-face context and F = 0 if the student participated in the virtual context. Finally, $\varepsilon$ represents the random error. We assume that the random terms $\varepsilon^i$ have independent normal distributions with mean zero and constant variance.

Besides, $\beta_M, \beta_T, \beta_P, \beta_{HA}, \beta_{LA}$ and $\beta_F$ are the regression coefficients corresponding to the variables M, T, P, HA, LA and F respectively and they were estimated through the OLS (Ordinary Least Squares) method.

This model was fitted using the function lm() in the programming language R version 4.1.0 [25].

**RESULTS**
**Comparison between contexts**

In our previous work [16], we compared the results of our KA model with the final grade that the students obtained. Now, we first compare the general results in both, face-to-face and virtual contexts.

We proposed in Eq. (1) that the final knowledge reached by a student on a given topic is mainly due to three contributing factors, the personal motivation (M), the influence of the teachers (T) and the influence of peers (P). In Fig. 3 we show the average values of the final grade of each group, <Kf>, together with the data obtained from the surveys carried out, in order to analyze and compare the differences observed with the change of context.

In Fig. 3(a) we show the average final grade <Kf> for each group of students (HA, AA and LA) and in both contexts. We observe again the differences we first noted in Fig. 1, related to how the performance of each group is modified with the change of context. For HA students, <Kf> increased during the virtual context while for AA and AA it decreased. In what follows, and to deepen the understanding of what is observed, we will analyze what was obtained for the three contributing factors, plotted in panels (b), (c) and (d) of Fig 3.

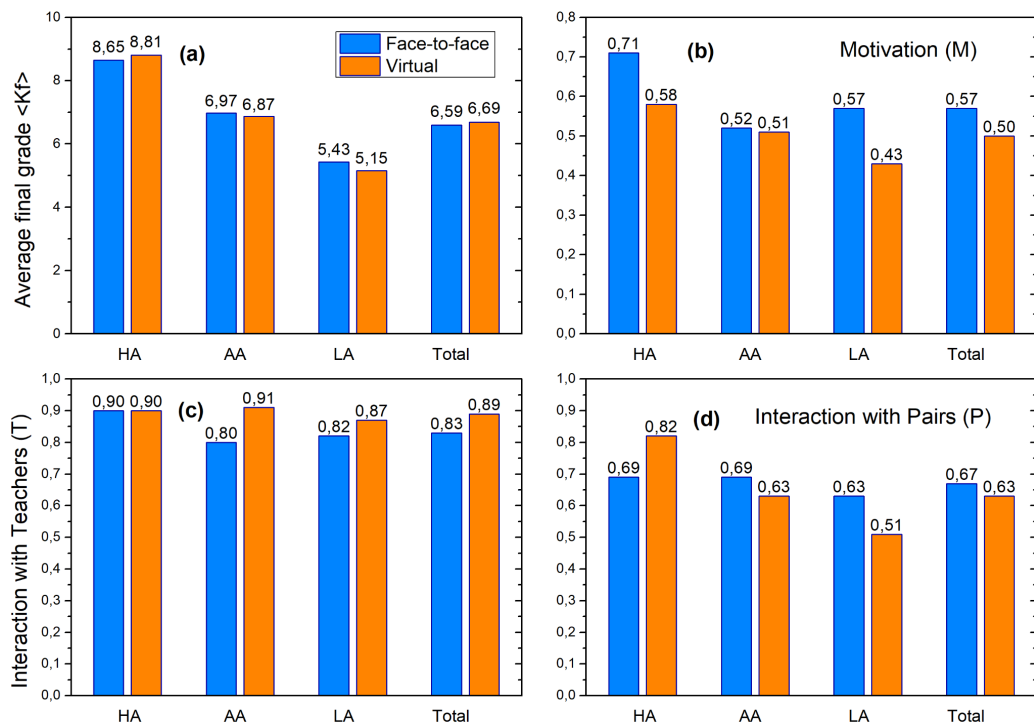

Fig 3: Numerical values associated with different quantities discriminated according to context, face-to-face (blue) and virtual (orange), averaged for each group of students HA, AA and LA: (a) Final grade (understood as final knowledge) <Kf>, (b) motivation M, (c) interaction with teachers T and (d) interaction with peers P.

The values of motivation presented in Fig. 3b reflects a widely studied aspect of the psychological impact of the pandemic on students [11,17]. Our results clearly report the impact of the virtual context on the motivation of students, no matter the group they belong to. This fact should in itself be an alarm to build policies to support the mental health and educational success of the students at all times. If motivation dropped notably in the new virtual context, and the final knowledge is considered as the sum of several factors that contribute to the acquisition of this knowledge, then the way of interacting with peers and teachers also had to change.

Fig 3c gives us information about the teacher's contribution from the students' perspective. Note that for the HA group it has the same weight in both contexts (face-to-face and virtual), while for the AA and LA groups the interaction with teachers increased in the virtual context. Generally, the teacher acts as an intermediary between the activities carried out by the students in order to assimilate the new knowledge and in this new context their "presence" and support was fundamental for many students.

Finally, in Fig 3d, we can see the differences in the interaction between peers for each group of students, another issue that was affected during the pandemic. We can see that HA's enriched the study in groups in the virtual context in contrast to the other groups of students. We also found that the structure of the emerging contact network from peer interaction presents very different characteristics in both contexts. More details about this aspect of the problem are presented in the next subsection.

The situation observed in the interaction of peers (Fig. 3d) is the one that most reflects the behavior of the general performance (Fig. 3a), however the trend is attenuated due to what is observed in Figs. 3b and 3c.

The aforementioned results can be summarized in Table 3 where we show the relative changes between both contexts. This quantity expresses what it was observed in Fig. 3 with the raw data obtained in the surveys: A strong decrease in the motivation term for all groups of students, and different trends in the way of interacting with peers and with teachers depending on the group to which the students belong.

Table 3
Relative changes in % for the quantities involved in the KA model between both contexts, calculated as Δ/reference, where Δ is the subtraction between the values in the virtual and face-to-face contexts, and the reference is the value in face-to-face context.

|  | relative change (%) | | | |
| --- | --- | --- | --- | --- |
| Quantities | HA | AA | LA | Total |
| Kf | 1.8 | -1.5 | -5.1 | 1.5 |
| M | -17.5 | -2.2 | -23.7 | -12.3 |
| T | -0.1 | 12.8 | 5.9 | 7.1 |
| P | 17.8 | -8.5 | -19.7 | -5.8 |

**Networks of peer interactions**
The analysis carried out around Fig. 3 indicates that the change in physical context modified the way in which students interact with each other. Furthermore, in this area data was collected in different ways depending on the context.

In the face-to-face context, the observations in the classroom were made in situ, with photographic records and paper surveys. During the virtual context, the surveys were digital using Google tools as mentioned above. In the latter case, no observations could be made, so the students were asked how their interaction with the group was and with whom they specifically interacted. This fact could result in a lack of information for this context. However, we observed that this was not the case, since although the information collected in both cases is not completely comparable, they suggest a change in behavior in the relationship between peers.

Table 4 expresses the number of students who were observed grouped or isolated during the face-to-face classes. Likewise, for the virtual case, the number of students who affirmed to study or not in a group is reported. We find that the percentage of isolated students decreased from 37% to 26% with the change of context. Interestingly, the increase in interaction between students in the virtual context was observed to a greater or lesser extent for the three groups.

Table 4. Number of students who were grouped or claimed to be in a group and the isolated students for each context.

| Context | Isolated students | | | In groups | | |
|---|---|---|---|---|---|---|
| | HA | AA | LA | HA | AA | LA |
| Face-to-face | 3 | 12 | 15 | 10 | 22 | 19 |
| Virtual | 2 | 11 | 11 | 14 | 30 | 22 |

To deepen the understanding of how students modified their way of interacting, we draw in Fig. 4(a) the network that represents the students before confinement (face-to-face context) for N = 81. As we said, the data was obtained from direct observations in the classroom, where the nodes represent the students (divided in the HA, AA and LA groups) and the links their interactions. Note that here we use double bonds, indicating a reciprocal interaction.

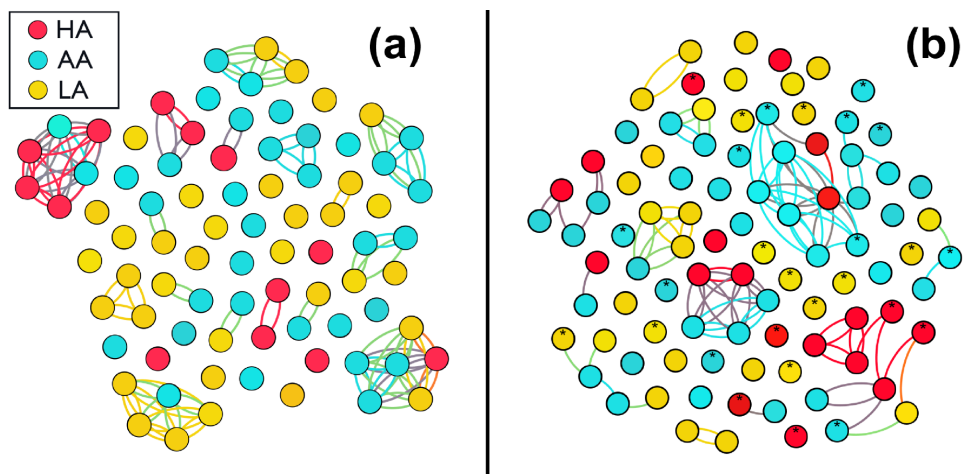

Fig. 4. Peer interaction of different groups of students in both contexts. (a) Network scheme from classroom observations in face-to-face context where the links are reciprocal interactions. (b) Network scheme from data obtained through surveys in virtual context. The links can be or not be reciprocal interactions. The nodes marked with an asterisk represent students who claimed to interact with students from another section who did not participate in this study.

Besides, in Fig. 4(b) we show the network that describes the students in virtual context for N = 92. The data come from the surveys carried out, and again the nodes represent the students divided in the groups HA, AA and LA. We use links to represent their interactions, although now they are double or single, as the responses to the surveys given by the students may or may not be reciprocal. Moreover, the nodes marked with an asterisk represent students who claimed to interact with students from another section who did not participate in this study.

A comparison between both networks indicates some similarities, such as the presence of highly connected clusters, as well as isolated students. However, the network corresponding to the virtual context has nodes that connect two different clusters, acting as ``bridges". This was not observed in the face-to-face context and could mean a new form of relationship between students.

**Measure of the relevance of the terms that influence the KAP**

A way to validate the model presented in Eq. (1) is to analyze the relevance of the terms that compose it. In our previous work [16] we did it by adding coefficients to each factor of the KA model. These coefficients could be interpreted as the relative weight that each term in Eq. (1) has, and were chosen so that the average value calculated with the model for each group is as close as possible to the average value of the actual final grades obtained.

Now we choose two different and complementary approaches to find the weight of each term of Eq. (1): Artificial Neural Networks (ANN) and a Multiple Linear Regression Method (MLR).

**ANN Approach**
Using SLP ANN we were able to get the coefficients of Eq. (1) as shown in Table 5. The values obtained indicate a generality in the weights of each factor that participates in the knowledge acquisition process, with greater prominence in the term associated with the interaction with teachers ($\beta_T^X$) for all groups. This is clearly observed in Fig. 5, where a graphical representation of the coefficients is shown. These results provide robustness to the model, as the weight that each factor has in the process we are describing is independent of the context analyzed. Moreover, and as will be clear soon, the ANN predicted similar coefficients to those obtained by MLR (see next section).

Table 5
The coefficients $\beta_M^X$, $\beta_T^X$ and $\beta_P^X$ obtained by ANN.

|  | Face-to-face context | | | Virtual context | | |
|---|---|---|---|---|---|---|
|  | HA | AA | LA | HA | AA | LA |
| $\beta_M^X$ | 3,2 | 3,1 | 2,2 | 3,3 | 3,1 | 2,2 |
| $\beta_T^X$ | 4,2 | 3,9 | 3,3 | 4,3 | 3,9 | 3,7 |
| $\beta_P^X$ | 3,2 | 2,7 | 1,8 | 3,2 | 2,7 | 1,3 |

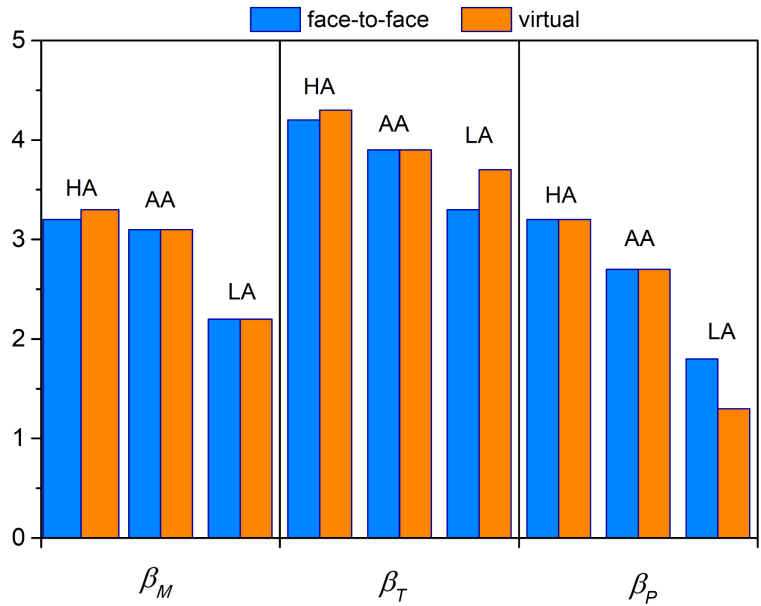

Fig 5: Coefficients $\beta_M^X$, $\beta_T^X$ and $\beta_P^X$ for each group of students HA, AA and LA obtained with the ANN approach. For each context these coefficients preserve generality with greater importance in the term associated with the interaction with teachers ($\beta_T^X$).

Finally, in Fig. 6 we present a comparison between the final grade for each student and the final knowledge obtained from Eq. (1) (KA model) with the coefficients obtained with the ANN approach. The global behavior of the KA model follows the general trend of the data. The observed dispersion is due to the presence of particular cases, whose complete evolution is not captured by the model. In our previous work [16] we made an analysis of some particular cases like these.

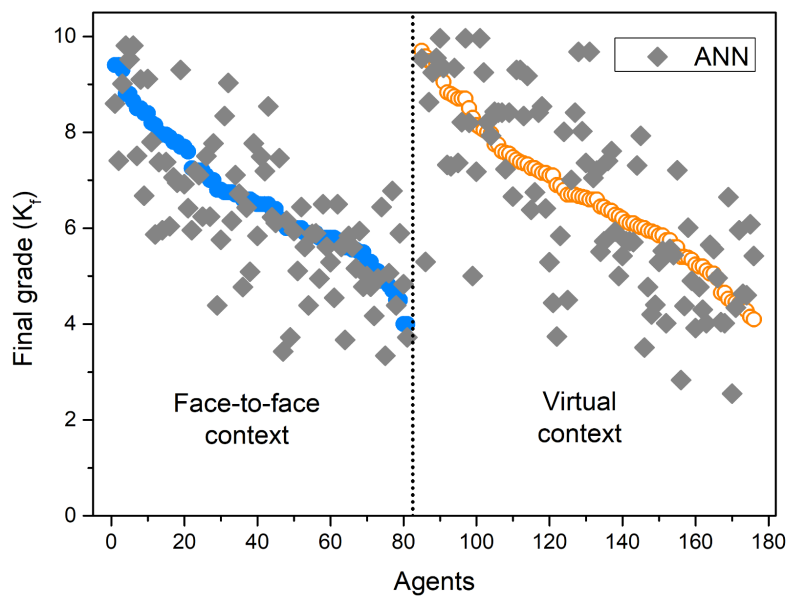

Fig. 6. Final knowledge comparison between the actual final grade obtained for each student in both contexts (circles) and the KA Model of Eq. (1) using the coefficients obtained with the ANN approach (diamonds).

**MLR Approach**

We use the Multiple Linear Regression Method in order to find the weights of each contributing factor of the KA model, and compare them with the ones obtained in the previous section. The results are shown in Table 6, where we express the values for $\beta$, SE and p-value for the terms of the Eq. (3).

Table 6: Regression results of multiple linear model of Eq. (3)

|  | $\beta$ | SE | p-value |
|---|---|---|---|
| **M** | 2.2776 | 0.3768 | 1.00e-08 |
| **T** | 4.6893 | 0.3063 | < 2e-16 |
| **P** | 1.4660 | 0.2308 | 12.09e-09 |
| **Student group according to their final achievements Kf** (Reference -> AA) | | | |
| **HA** | 1.9131 | 0.2664 | 2.43e-11 |
| **LA** | -1.0585 | 0.2136 | 1.81e-06 |
| **Context** (Reference -> Virtual context) | | | |
| **Face-to-face context** | 0.7424 | 0.1902 | 0.000139 |
| | | | |
| | | **Adjusted R-squared** | 0.9664 |
| | | **p-value** | <2.2e-16 |

Note: the lower limit of the values for T was modified to coincide with the scales of the rest of the variables.

The p-values obtained show that all beta regression coefficients are statistically significant. Assumptions of linearity, independence, homoscedasticity and normality were checked, as well as the presence of influential values.

We find that the term with the highest weighting is the one related to the interaction with teachers ($\beta_T^x$), in accordance with what was obtained with the ANN approach.

At last, we show in Fig. 7 a comparison between the final grade for each student and the final knowledge of Eq. (1) (KA model) with the coefficients obtained with the MLR approach. Again, the $K_f$ obtained with the model behaves similarly to the data. It should be noted the similarity of the result obtained in Figs. 5 and 6 with that shown in Fig. 2 of [16]. In the present work, the adjustment of the weights that gave rise to both figures was carried out in a more appropriate way than in that paper, where the coefficients of each term were chosen exploratory.

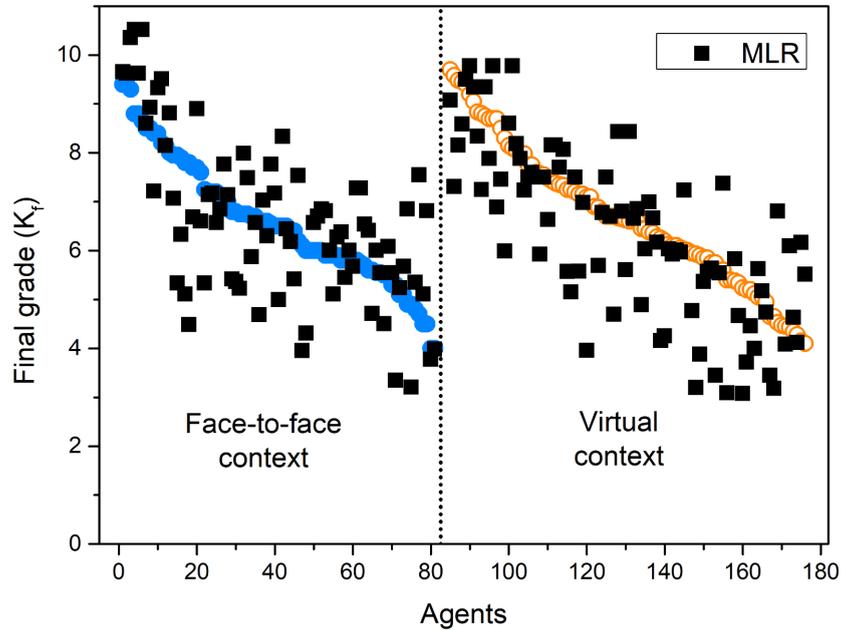

Fig. 7. Final knowledge comparison between the actual final grade obtained for each student in both contexts (circles) and the KA Model of Eq. (1) using the coefficients obtained with the MLR approach (squares).

**Discussion**

We analyze various quantities that participate in the knowledge acquisition process in face-to-face and virtual contexts for a specific study case. Our investigation spanned two years and involved 173 students, observing the evolution of their learning process for each particular context.

The raw data in Fig. 1 show that the final grade of the students in both contexts presented differences. Specifically, the grades of the students with high performance (HA) were better in virtual context than in face-to-face context. The opposite is seen with low-achieving students (LA), while the intermediate performance group (AA) did not show differences.

Inspired by a previous work, we wanted to assess whether the changes observed in academic performance could be understood from a model that incorporated the main factors that contribute to the knowledge acquisition process. In order to do this, we developed an analytical data-based model, that we call the KA model, consisting of a series of surveys that are contrasted with information on academic performance. These surveys were carried out 3 times during each semester, and reflected the feelings of the students during their learning process that were reflected in some way in their performance.

The numerical values assigned to the survey responses shown in Fig. 3 reinforce accepted ideas related to the importance of motivation in the learning process: the switch to virtual context caused a decrease in motivation to learn and this is strongly reflected in the performance of the LA students. This fact should alert the educational community and especially those responsible for building support mechanisms for the mental health of students. Furthermore, we observe that the new context generates a change in the way students interact with their peers and teachers. In particular, the HA students did not modify the interaction with the teachers (maintaining high values in both contexts) while they strengthened the study in groups in the virtual context, unlike the rest of the groups. For AA and LA students, interaction with teachers increased in the virtual context, and this result highlights the importance of the teacher's role as a consultant and as fundamental support for students.

We also find that the structure of the network of contacts that is formed between peers in both contexts presents some common characteristics, as well as some interesting differences, as we saw in Fig. 4. Among the first is that both networks have highly connected clusters, as well as a significant number of completely isolated students.The virtual context network, however, shows a characteristic not observed in the other network: the presence of individuals who interact with one or more students from different clusters. These individuals act as bridges between students who otherwise would not be connected. These structures could be reflecting a new form of relationship between students that occurs more easily in the virtual context. Nevertheless, we are aware that this analysis requires a more detailed investigation that is beyond the scope of this work with the data that we currently have.

Related with the previous analysis is the fact that, although the equation in the KA model is linear, the term of peers can be interpreted as an effective version of a real non-linear interaction. This term in itself adds complexity to the model since group interaction does not obey "linear" rules. However, the simplification made in the KA model remains valid in light of the results obtained in [16] and are in line with the idea that the learning process is not limited to the interactive behavior of individual teachers and students, but should be understood in terms of collaborative behavior [26].

In order to validate the proposed KA model, we explored two methods to find out the relevance of the factors that we consider to be most important in the knowledge acquisition process: a standard Multiple Linear Regression Method and a Single Layer Perceptron, which is a particular type of Artificial Neural Network.

The results obtained with the neural network (Fig. 5) indicate that in both contexts the weights are similar. This result also shows that the raw results are those that speak of each context, as the data obtained in each situation are those that describe the particular reality that each group of students goes through. Moreover, both approaches indicate a greater relevance of the term of interaction with teachers. We were able to gather information from the teachers to support this fact and the sensation in the change of interaction was also commented on by them (see S1 File). The knowledge acquisition process comes hand in hand with the importance of the interaction with teachers, and the literalness of their presence in the accompaniment during learning.

The comparisons of Figs. 6 and 7 between the raw data and the results obtained with the KA model indicate that the general behavior of individuals can be suitably described with Eq. (1), which is simply the sum of the relative contributions of each of the proposed factors: personal motivation, interaction with pears and influence of teachers. The robustness of the coefficients obtained with the two approaches also indicates that the information collected in the surveys and observations was sufficient to construct an adequate representation of the process. We are aware that this simplification leaves out a huge number of variables that are integrated to give rise to the unique process that each person experiences. But we believe that the results obtained allow us to validate our choice of factors as the main contributions common to all individuals.

Finally, we discuss some considerations on the scope and limitations of this work.

One is that we must not lose sight of the fact that the change in the specific physical context brought with it a change in the evaluation criteria. Actually, this aspect was addressed in the teacher interviews that we summarize in the Supplementary Material (S1 File). As $K_f$ is a hard data (the final grade obtained in the course), it would be more appropriate to build new models that consider these data in a more comprehensive way, taking into account the challenges that arose due to the change in this educational context.

Another important issue that is absent from the KA model is the personal context of the students and their available resources. The reason why it was not included is because we had no survey done on these topics in the face-to-face period, so it was not possible to compare both contexts. However, in the Supplementary Material (S1 File) we include additional information regarding this subject obtained from the surveys carried out in the virtual context. When asking the students for their feelings regarding confinement, the responses were varied but reluctance was reflected in more than half of the responses. This coincides with our observation about the lack of motivation (see Fig. 3b). The emotional stress, widely discussed in this context, goes beyond the academic environment and it was an important characteristic that we tried to capture with our research. Moreover, we found some relevant differences between the students of the different groups, which could influence their performance. Among them, a third of the students belonging to the LA group said they had a poor Internet connection in contrast to the HA group in which this situation occurred for a sixth of the students. More importantly, 13% of students belonging to the LA group did not have a laptop computer and 30% did not have an adequate study space. The importance of recognizing inequalities lies in making visible the urgent need to build university policies to improve this situation.

The last comment is that our study of the virtual context was carried out during the first year of the pandemic, so the results obtained could be strongly influenced by the transition between both contexts. However, we believe they are valuable in themselves and can serve to deepen the understanding of the complex process of learning.

**Supporting information**
S1 File. Additional information obtained from student and teacher surveys (is available upon request from the authors). (PDF)